\documentclass[]{jfm_arxiv}

\usepackage{graphicx}
\usepackage{xcolor,soul}
\usepackage{mathtools}
\usepackage{newtxtext}
\usepackage{newtxmath}
\usepackage{natbib}
\usepackage{hyperref}
\usepackage{placeins}
\usepackage{cases}
\usepackage{overpic}
\usepackage{siunitx}
\usepackage{textcase}

\hypersetup{
    colorlinks = true,
    urlcolor   = blue,
    citecolor  = black,
}

\newcommand{\RomanNumeralCaps}[1]

\title{Morphology of mucus films in lung airways: secretion and ciliary evacuation
}

\author{Swarnaditya Hazra\aff{1}
 \and Jason R. Picardo\aff{1} \corresp{\email{picardo@iitb.ac.in}}}

\affiliation{\aff{1}Department of Chemical Engineering, Indian Institute of Technology Bombay, Mumbai 400076, India}

\begin{document}

\maketitle

\begin{abstract}
Lung airways are lined by a film of mucus which protects the epithelium from inhaled particles. To maintain a uniform coating, the mucus that is secreted into airways must be distributed into a film by wall-attached cilia, which constantly convey mucus along the airway. 
However, the film's natural tendency is to accumulate into humps and plugs, due to the Rayleigh-Plateau instability. To understand the behaviour of the film amidst these competing factors, we perform simulations of an idealized tubular airway using a reduced-order thin-film model. The axial boundaries are nonperiodic, allowing for cilia-driven inflow and outflow; a tangential velocity along the tubular wall models ciliary transport, while a localized source at the wall accounts for secretion. On increasing the mucus input rate, we find three distinct film morphologies: (i) uniform flat films; (ii) nonuniform films that are composed of travelling unduloid-shaped humps, separated by mucus-depleted zones; and (iii) films that form an occluding plug. The flat-film regime, absent in closed periodic domains, emerges as a consequence of the convective nature of the Rayleigh-Plateau instability in the presence of ciliary transport. Higher secretion rates increase the mean film-thickness and induce a convective-to-absolute transition, which manifests in the appearance of travelling humps.
In the plug-forming regime, capillary forces dominate and drive the incoming mucus into a single hump, which resists ciliary translation and remains near the mucus source.
Our results show how low-level baseline secretion sustains a protective, uniform mucus film, and how hypersecretion---stimulated, for example, by inhaled allergens---produces mucus plugs.
\end{abstract}


\section{Introduction}

\begin{figure}
\centering
 \includegraphics[width=0.86\textwidth]{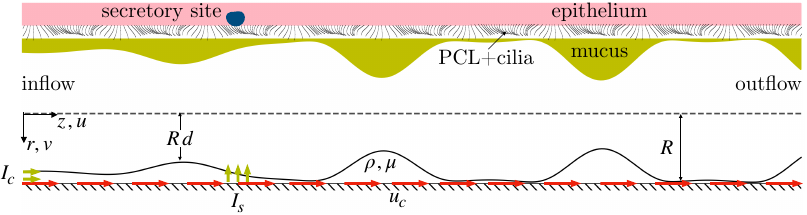}
\caption{\label{fig:schematic} Illustration of a mucus-lined ciliated airway with a mucus-secretion site (top half) and schematic of the corresponding model (bottom half). The airway, of length $L$, has open boundaries in the axial direction through which mucus flows in and out of the domain.}
\end{figure}

Mucus is important for the healthy functioning of the lungs. Present as a film along the inner walls of conducting airways, mucus acts as a barrier against inhaled particles, including allergens and pathogens \citep{Boucher2006}. 
Particles that deposit on the mucus film are conveyed out of the lungs, along with the mucus, by a carpet of wall-attached cilia. The cilia are submerged in a sub-layer of watery periciliary liquid (PCL) and their tips penetrate into the overlying mucus film (figure~\ref{fig:schematic}, top half); synchronous ciliary beating, in the form of a metachronal wave, produces a net transport of mucus and constantly evacuates mucus from the airways \citep{Sleigh1988}.

The loss of mucus by ciliary transport is balanced by secretion into the airways, from submucosal glands and goblet cells present in the epithelium of the airway walls~\citep{Rogers2033-goblet}. Secretion occurs in two modes: slow but sustained baseline-secretion, and rapid but relatively short-lived stimulated-secretion~\citep{William2015-baseline,Dickey2018-secretion}. The former is thought to be the normal mode of secretion that maintains a uniform mucus film, which provides routine protection from inhaled particles; the latter is triggered by various stimuli including allergens and pathogens. Hypersecretion of mucus is known to produce plugs which occlude the airway and obstruct airflow~\citep{Levy2014}. It has been hypothesized that stimulated hypersecretion acts as a defence against invading microorganisms like helminths (parasitic worms)---the formation of a mucus plug might trap a helminth in the conducting airways and prevent it from migrating deep into the lungs~\citep{Dickey2018-secretion}. This mechanism goes awry in conditions like allergic asthma, wherein mucus plugs block several airways and cause respiratory distress~\citep{rogers2004airway}. 

The formation of plugs is driven by the surface-tension of the mucus-air interface \citep{Levy2014,Romano2019viscous}. In the absence of other physical effects, an initially uniform annular-film spontaneously undulates due to the Rayleigh-Plateau (or capillary) instability and ultimately adopts one of two equilibrium configurations \citep{johnson1991}: (i) a non-occluding annular hump, called an unduloid, or (ii) an occluding liquid-bridge or plug. Unduloids exist only up to a critical volume beyond which they give way to plugs~\citep{everett1972model}. The fate of a film in a long tube depends on how the liquid is dynamically partitioned by the capillary instability, as well as on the subsequent nonlinear dynamics \citep{hazra2025probabilistic}; nonetheless, sufficiently thick films always form plugs. It is not surprising then that stimulated hypersecretion produces plugs. 

In contrast, we hitherto lack a physical explanation of the uniform-film configuration, which is supposed to prevail in healthy airways with low-level baseline secretion. The capillary instability is unavoidable and so an initially uniform film, no matter how thin, will accumulate into a series of unduloids, separated by mucus-depleted zones~\citep{lister2006capillary,hazra2025probabilistic}. The width of these zones is sizeable, with a minimum value of about 30\% of the tubes length; these depleted zones widen as the film thickens because capillary forces produce deeper but narrower unduloids~\citep{hazra-particles}. Hence, it appears that a uniform flat film is impossible and that a portion of the airway is inevitably left exposed to aerosol deposition~\citep{hazra-particles}. 

Here, we show that ciliary transport can maintain a uniform flat film by convecting interfacial perturbations out of the airway. However, to realize this effect of cilia, one must relax the assumption of periodic axial boundaries; indeed, in a periodic domain, ciliary transport has been shown to have no impact on the films' morphology \citep{hazra-particles}. Using an outflow boundary condition allows the evacuating action of ciliary transport to alter the film's dynamics. The consequent loss of mucus must be balanced either by cilia-driven inflow at the entrance of the airway or by secretion within the airway; both situations are considered here and shown to be closely related.

Our work uncovers three distinct film morphologies; on increasing the rate of mucus influx relative to ciliary transport, the film transitions from a steady and flat configuration to a dynamic state, composed of travelling humps, and ultimately to a plugged state. We also find that open-domain films, maintained by secretion and ciliary evacuation, plug more readily---the minimum film thickness required for plug formation is less than that in a closed periodic-domain.

\section{Mathematical model}
The lung has 24 generations of branching airways, beginning with the trachea (generation 0) and terminating at the alveoli. Our study focuses on the middle conducting airways (generations 10-16), which are mucus-lined, ciliated, and relatively-rigid, i.e., not prone to capillary collapse like the terminal airways~\citep{heil2008mechanics}. 
Gravity is negligible in these small airways, and the airflow is laminar~\citep{pedley1977pulmonary} and too weak to alter the distribution of mucus under normal breathing conditions \citep{Dietze2015,hazra-particles}. The combination of weak airflow and a relatively-high mucus volume-fraction (compared to the large upper airways) makes the middle airways particularly prone to occlusion by mucus plugs.  

To facilitate simulations on long domains with non-periodic boundaries, we make several simplifying assumptions. The flow of air is ignored and the mucus is treated as a viscous Newtonian fluid. The sublayer of PCL and cilia is replaced by a non-penetrating boundary---a mesh of cross-linked polymers prevents mucus from entering the PCL sublayer under healthy conditions \citep{button2012}. The effect of ciliary transport is retained, in a coarse-grained manner, by imposing a tangential velocity $u_c$ at the wall. Previous studies have successfully used this approach to reproduce experimental observations of large-scale mucociliary transport~\citep{vasquez2016modeling,Bottier2017model}. To mimic the metachronal wave of cilia-beating, an asymmetric travelling wave should be used for $u_c$; however, unless the mucus is abnormally viscoelastic (as occurs under diseased conditions), its transport is determined entirely by the mean of $u_c$ \citep{Dietze2023mucociliary}. Indeed, we have shown in earlier work that the dynamics of a viscous mucus film remains the same when a travelling wave form for $u_c$ is replaced by a constant velocity that equals its mean \citep{hazra-particles}. We therefore proceed with a constant value for $u_c$. Typical mucociliary transport velocities, in healthy airways, range from $25$ to $60$ \si{\mu m.s^{-1}}, and we choose $u_c$ accordingly.

A schematic of our idealized airway is presented in the bottom half of figure~\ref{fig:schematic}. An axisymmetric mucus film of viscosity $\mu$ and density $\rho$ lines the inner wall of a cylindrical tube of radius $R$ and length $L$. The free surface of mucus, which has surface tension $\gamma$, is located at $r = d(z,t)$. 
All variables are scaled with $R$, $\tau_{RP}$, and $U = R/\tau_{RP}$, 
where the time scale $\tau_{RP} = ({\mu R}/{\gamma}) 24 (d^{*}_0)^5[(1-d^{*}_0)^2(1-(d^{*}_0)^2)]^{-1}$ is the inertialess approximation to the inverse growth rate of the fastest-growing Rayleigh-Plateau mode~\citep{johnson1991,Dietze2015}, for a film of initial thickness $1-d^{*}_0 = 0.03$ (this value lies in the middle of the range of thicknesses considered here).  The dimensionless flow problem is governed by the capillary number $Ca ={\gamma \tau_{RP} }/{\mu R }$ and the Reynolds number $Re = {\rho R^2}/{\mu \tau_{RP}}$. Using physiologically-relevant parameter values, $R = 400$ \si{\mu m}, $\mu = 0.01$ \si{Pa.s}, $\rho=1000$ \si{kg.m^{-3}}, and $\gamma = 0.05$ \si{N.m^{-1}} \citep{hazra-particles}, we obtain $Ca = 3.8 \times 10^{5}$ and $Re = 5 \times 10^{-4}$. Clearly, inertia is negligible in the mucus film ($Re = 2 \times 10^{-3}$ if a typical value of the ciliary transport velocity, 60 \si{\mu m.s^{-1}}, is used as the velocity scale), and so we consider Stokes flow throughout this work .

Ciliary transport will evacuate mucus from the airway of interest, while simultaneously bringing in mucus from the upstream airway. We therefore implement a free outflow boundary condition at the right-end of the airway, while imposing an input flow rate $2\pi I_c$ at the left-end (figure~\ref{fig:schematic}). This scenario corresponds to mucus being secreted into a different, upstream airway and being transported into the airway of interest by cilia. We also study the scenario in which secretion occurs directly inside the airway of interest; this case is modelled by having a localized source term 
at the wall which injects mucus at a rate of $2 \pi I_s$. 
The first scenario admits a flat-film base state, 
whereas the second scenario necessarily has a spatially-varying film thickness. Nonetheless, the dynamics in the second scenario turn out to be closely related to that in the first.


The flow of mucus is described by the thin-film model of \citet{Dietze2015}, namely, the second-order Weighted Residual Integral Boundary Layer (WRIBL) model. On neglecting air flow and inertia, the model consists of the following equations for the interface position $d(z,t)$ and the flow rate $2\pi Q(z,t)$: 
\begin{equation}\label{mass_balance}
d \partial_t d= \partial_z Q+s,
\end{equation}
\vspace{-2. em}
\begin{multline}\label{wribl_q_eq_pas}
 {M}_m{\partial_z^2Q}+{K}_{m}\partial_z Q\partial_z {d}
 +{L}_{m}Q{\partial_z^2 {d}}+{J}_{m}Q (\partial_z {d})^2 + Q=\\ {u_c{(1-d^2)}/{2} } -Ca(\partial_z\kappa+\partial_z\kappa_p)  {N_m}- {J}_{c}u_{c}(\partial_z {d})^2 +\left({L}_{c}u_{c} \right){\partial_z^2 {d}}.
\end{multline}
The source $s(z)$, in \eqref{mass_balance}, is zero everywhere except in the region $z_s\leq z\leq z_s+\ell$ where it takes a constant value $I_s/\ell$, so that $\int^{L}_{0} s dz = I_s$; this localized source models secretion from a region of width $\ell$, at a distance $z_s$ from the inlet of the airway (in simulations, we smoothen the square pulse using $\tanh$ functions). The coefficients of \eqref{wribl_q_eq_pas}, $J_c$, $L_c$, $M_m$, $K_m$, etc., are functions of $d$ alone. The corresponding expressions are provided in a Jupyter notebook (see the caption of figure~\ref{fig:saddle}). 
A derivation of the WRIBL equations, using computer algebra, is given by \citet{HazraWRIBL}.

The WRIBL model has been extensively validated against direct numerical simulations of annular films~\citep{Dietze2015}. Even in the inertialess limit, the WRIBL model retains key advantages over traditional lubrication theory, because it consistently accounts for the effects of longitudinal viscous stresses and nonlinear interfacial curvature~\citep{Dietze2015}. The latter is essential for capturing the formation of plugs. The second-order approximation for the curvature used in \eqref{wribl_q_eq_pas} is \(\kappa = d^{-1} - \partial_{zz} d - (\partial_z d)^2 (2d)^{-1}\). This expression yields a maximum unduloid volume of 1.73 $\pi R^3$, in good agreement with the exact full-curvature result of 1.74 $\pi R^3$; the shapes of stable unduloids are also well predicted \citep{hazra-particles}. 

To simulate the dynamics of the film with mucus-depleted zones, we introduce a precursor film \citep{oron1997long} by adding a disjoining pressure $\kappa_p = {h^6_0}{(1-d)^{-9}}$ in \eqref{wribl_q_eq_pas}. Setting $h_0 = 1\times10^{-4}$ yields a precursor film thickness of about $ 0.01$; hence, regions of the wall where $1-d \approx 0.01$ correspond to dried-out zones.

We set the size of the domain, $L$, in terms of the wavelength $\Lambda_{RP}$ of the fastest-growing Rayleigh-Plateau instability mode. This wavelength dominates the pattern of humps and depleted zones that emerge from an initially-quiescent and unbounded annular film~\citep{hazra2025probabilistic}. A good approximation to this wavelength is provided by the inviscid analysis of \citet{rayleigh1892xvi}: $\Lambda_{RP} \approx 2\sqrt{2}\pi d_0$, where $d_0$ is the radial position of the flat-film base state \citep{Dietze2015}. For the films considered here, $d_0 \gtrsim 0.95$, and so $\Lambda_{RP} \approx 2\sqrt{2}\pi = \Lambda$. With the parameter values used here, $\Lambda R \sim 3$ \si{mm}; in comparison, an airway's length is about 5 - 10\si{mm} \citep{Sleigh1988}. So, airways are long enough for multiple mucus humps to appear. We therefore use $L = 4\Lambda$ in our simulations. In cases with secretion, we increase the domain length to $L = 5 \Lambda$ in order to better understand the effects of secretion.


Regarding boundary conditions, we note that \eqref{wribl_q_eq_pas} requires three conditions on $d$ and two on $Q$. At the inlet, we pin the interface position and specify the incoming flow rate of mucus:
\begin{align}\label{left_bc}
 d = d_0, \quad Q = I_c , \quad at \quad z = 0.
\end{align}
In the absence of secretion, this inlet condition allows for a flat-film base steady state in which $d=d_0$ and $Q = I_c =  u_c{(1-d_{0}^2)}/{2}$ (to satisfy \eqref{wribl_q_eq_pas}). At the outlet, we impose soft boundary conditions which permit mucus to exit the domain:
\begin{align}\label{right_bc}
\partial_z d = 0,\quad \partial^3_z d = 0, \quad \partial^2_zQ =0 , \quad at \quad z = L.
\end{align}
These exit conditions alter the shape of the interface near the exit, especially when a mucus hump leaves the domain. We therefore extend the physical domain by $L_\Delta$ so that the computational domain has a length $L+L_\Delta$. We find that $L_\Delta = 3\Lambda$ is sufficient to ensure that the film dynamics within the physical domain $(0,L)$ are unaffected by the outlet boundary. 

As in previous work \citep{hazra-particles,hazra2025probabilistic}, we numerically solve \eqref{mass_balance} and \eqref{wribl_q_eq_pas} using a second-order finite-difference scheme to discretize space; ghost nodes are used to implement the boundary conditions. Time integration is performed using the adaptive time-stepping solver solve{\textunderscore}ivp from the SciPy library in Python~ \citep{scipy}. A fully-implicit and variable-order scheme is implemented based on backward difference formula (BDF). The computations are eased by \eqref{wribl_q_eq_pas} being linear in $Q$ when $d$ is known: the calculation of $Q$ on the grid, given a guess for $d$, amounts to solving a linear system of equations (we use the linalg.solve routine of NumPy~\citep{Numpy2020}). The iterative calculation of $d$ and $Q$, within each update of the fully-implicit time-stepper, benefits from this linearity in $Q$ of the noninertial WRIBL model. 
All our simulations are initialized by perturbing the flat film base state, using the wavelength of the fastest-growing Rayleigh-Plateau mode; to ensure compatibility with the boundary conditions, we initialize the internal grid points using $d_0 + 10^{-3}(1-d_0)\,\sin(2\pi z/\Lambda)$ and then calculate the initial values for the boundary nodes using the boundary conditions.  

\section{Morphology of transported films}

\begin{figure}
\centering
 \includegraphics[width=0.88\textwidth]{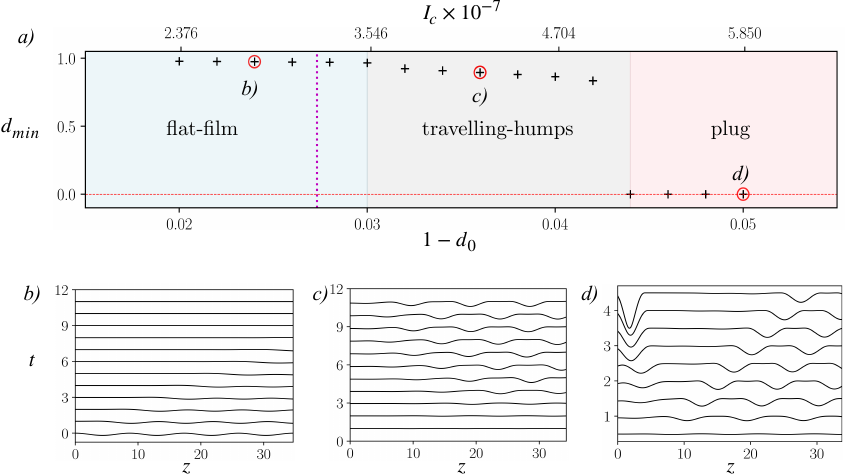}
\caption{\label{fig:regime_map} (\textit{a}) Map demarcating the dynamical regimes 
encountered on varying the mucus input flow rate $I_c$, which in turn sets the thickness $1-d_0$ of the flat base-state (see the top and bottom horizontal axes of panel (a)). The vertical axis shows the minimum position attained by the interface, $d_{min}$. Illustrations of the film's evolution in the three regimes are presented in panels (b-c). Here $u_c U = 60$ \si{\mu m.s^{-1}}; analogous results for $u_c U = 40$ \si{\mu m.s^{-1}} are presented in the \href{https://bighome.iitb.ac.in/index.php/s/yeijAoz47zg7yZk}{ supplementary material}.
}
\end{figure}

We begin with the first scenario in which secretion occurs not within the airway of interest ($I_s = 0$) but in a different upstream airway; ciliary transport then draws mucus into the airway of interest at a flow rate $I_c$. Fixing the ciliary velocity $u_c$ and increasing the input rate $I_c$ yields films of increasing mean-thickness; indeed, the thicknesses $1-d_0$ of the corresponding flat-film steady states increase with $I_c$ since $u_c{(1-d_{0}^2)}/{2} = I_c$. These states are susceptible to the Rayleigh-Plateau instability, however, and the consequent dynamics fall into three distinct regimes of behaviour. These regimes are mapped out as blue, gray, and red regions in figure~\ref{fig:regime_map}(a), for a ciliary transport velocity of $u_c U = 60$ \si{\mu m.s^{-1}}; the figure presents the minimum position realized by the interface throughout its evolution ($d_{min}$), for a range of simulations with increasing inlet flow rates $I_c$ (top axis) and hence increasing base-thicknesses $1-d_0$ (bottom axis). A value of $d_{min} > 0$ implies an open airway, while $d_{min} = 0$ marks the formation of an occluding mucus plug. The typical evolution of the film in the three regimes is illustrated in figures~\ref{fig:regime_map}(b-d) (see movie 1, movie 2, and movie 3 of the \href{https://bighome.iitb.ac.in/index.php/s/yeijAoz47zg7yZk}{ supplementary material}). 

At small $I_c$ or small $1-d_0$ (blue region), the thin films initially develop undulations, but these are convected out of the domain to leave behind a flat film (figure~\ref{fig:regime_map}(b)). At moderate $I_c$ (gray region), the undulations persist and grow, so that the moderately-thick film organizes itself into a sequence of travelling unduloid-shaped humps separated by depleted zones (figure~\ref{fig:regime_map}(c)); each time a hump is convected out of the airway, a new one forms near the inlet. Finally, at large $I_c$ (red region), the humps that emerge from the thick film are so deep that capillary forces overcome ciliary transport; a hump that forms near the inlet is not convected away but rather accumulates the incoming mucus until it forms a plug (figure~\ref{fig:regime_map}(d)). Plug formation manifests as a singularity in thin film models: the corresponding hump exhibits runaway growth and $d$ approaches zero with increasing rapidity. We terminate our simulations at a small but finite value of $d$ with the understanding that the final outcome is a plug (which is represented in figure~\ref{fig:regime_map}(a) by setting $d_{min} = 0$).

Our calculations for other values of $u_cU$, in the range of 20 - 60
\si{\mu m.s^{-1}}, show the same three regimes. As a second example, the regime map for $u_c U = 40$
\si{\mu m.s^{-1}} is presented in the \href{https://bighome.iitb.ac.in/index.php/s/yeijAoz47zg7yZk}{ supplementary material}; it is entirely analogous to figure~\ref{fig:regime_map}(a). 

Unduloid-shaped humps and plugs are the only film morphologies that emerge in periodic domains, even after including ciliary transport~\citep{hazra-particles}. It is only when the periodic boundary conditions are replaced by inlet and outlet conditions that the flat-film emerges. Changing the boundary conditions replaces a closed domain with an open domain, which enables ciliary transport to sweep away the Rayleigh-Plateau-induced undulations. This situation is typical of an absolute-to-convective instability transition.


\section{Spatio-temporal stability analysis}

We now examine the stability of the base state, $d=d_0$ and $Q = u_c ({1-d^2_0})/{2}$, on an unbounded domain. Linearizing \eqref{mass_balance} and \eqref{wribl_q_eq_pas} and assuming normal-mode perturbations of the form $\exp[i(k z-\omega t)]$, we obtain the following dispersion relation:
\begin{equation}\label{dispersion}
  \omega = \frac{-1}{{d_0^{3} (1-k^{2} M_{m})}} \left[\mathrm{i}\, N_{m}k^{2} \left(1 - d_0^2k^2\right) +{ u_c k d_0^{2} \left\{ \left(\frac{1-d^2_0}{2}\right) k^{2} L_{m} + k^{2} L_{c} - d_0\right\}} \right]
\end{equation}

From a temporal stability perspective (complex growth rate $\omega = \omega_r + i\, \omega_i$ but real wavenumber $k$) the flat film is always unstable, since the growth rate $\omega_i$ is positive for small but non-zero $k$ ($M_m$ and $N_m$ in \eqref{dispersion} are negative). Indeed, ciliary transport ($u_c$) does not alter the temporal growth rate, but causes the modes to travel with a $k$-dependent wave speed $\omega_r/k$. Hence, a flat film is never realized on periodic domains~\citep{hazra-particles}, wherein the spatial propagation of disturbances is irrelevant.

\begin{figure}
\centering
 \includegraphics[width=0.9\textwidth]{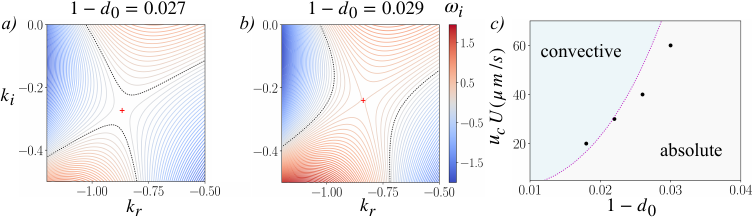}
\caption{\label{fig:saddle} (a-b) Contours of $\omega_i$ in the $k_r$-$k_i$ plane for convectively and absolutely unstable base films; the red "$+$" marker locates the saddle pinch point and the dotted lines correspond to the $\omega_i = 0$ contour. A Jupyter notebook that derives \eqref{dispersion} and plots panel (a) is available at \href{https://cocalc.com/share/public_paths/af07785cc665fccfb4c2939f0996ad2001ecb4b1}{https://cocalc.com/share/public{\textunderscore}paths/dade5dbc44da89bf9c905d19b83941294ff7914b}. (c) The convective-absolute boundary (pink-dashed) along with numerically-determined transition points (black circles) between the flat-film and travelling-hump regimes. }
\end{figure}

In contrast, the question of stability in a finite and open domain requires one to account for the possibility of perturbations leaving the domain. We therefore adopt the spatio-temporal approach, which accounts for the dispersive propagation of wavepackets produced by an impulse perturbation. Following standard methodology~\citep{Huerre2000Shear}, we allow both $\omega$ and $k$ to be complex, with $k = k_r + i\,k_i$; we then calculate the saddle point ($\omega_0,k_0$) at which $d\omega/dk = 0$ and examine the corresponding growth rate $\omega_{0i}$. As per the Briggs-Bers criterion, the wave-packet with zero group velocity---which remains inside the domain---will asymptotically grow (decay) if $\omega_{0i} > 0$ ($<0$); the system is then absolutely (convectively) unstable. 
We ensure that the saddle is a valid pinch point, by plotting contours of $\omega_i$ in the $k_r$-$k_i$ plane~\citep{rees2010effect}. Figures~\ref{fig:saddle}(a) and~\ref{fig:saddle}(b) present examples for convectively and absolutely unstable films. In both cases, we see that the saddle (red $+$ marker) is pinched between positive and negative spatial branches---the contours of $\omega_i$ which meet at the saddle move into the top and bottom halves of the $k_r$-$k_i$ plane as the value of $\omega_i$ is increased from $\omega_{0i}$. 

The spatio-temporal stability analysis shows that thin films (low $I_c$) are convectively unstable while thicker films (higher $I_c$) are absolutely unstable. The spatio-temporal nature of the instability changes at the critical base-state thickness for which $\omega_{0i} = 0$. 
The critical thickness for $u_c U = 60$ \si{\mu m.s^{-1}} is marked on figure~\ref{fig:regime_map}(a) by a vertical pink-dashed line; it appears to anticipate the transition from the flat-film regime to the travelling-hump regime in our open domain simulations.  
Figure~\ref{fig:saddle}(c) extends the comparison between the stability theory and the open domain simulations to other values of $u_c$. The convective-absolute boundary in the ($1-d_0,u_c$) parameter plane, computed via continuation~\citep{rees2010effect}, shows that the critical film-thickness increases with the ciliary velocity, in reasonable agreement with our open domain simulations---the solid circles represent the thickness up to which the flat film persists and beyond which travelling-humps emerge. Clearly, ciliary transport acts to convect perturbations out of the airway, thereby stabilizing flat films, provided they are sufficiently thin.





\section{Secretion within the airway}

We turn now to the scenario in which mucus is secreted inside the domain. Specifically, we fix the inlet flow rate to a small value $I_c^*= 2.38\times 10^{-7}$ that would produce a very thin, stable, flat film in the absence of secretion (see figure~\ref{fig:regime_map}(a)). We then turn on a secretory source, {located at $z_s = 2.3\Lambda=19.7$ and extending a short distance of $\ell = 0.05\Lambda = 0.4$}. In the absence of the Rayleigh-Plateau instability, the additional mucus that enters the domain at a rate $2 \pi I_s$ would be transported downstream. So, we expect a base steady-state in which the film is flat both upstream and downstream of the secretion zone but undergoes an increase in thickness from $1-d_{up}$ to $1-d_{dn}$ across the secretion zone (where $1-d_{up}^2 = 2I_c^*/u_c$ and $1-d_{dn}^2 = 2(I_c^*+I_s)/u_c$). If the downstream portion of the film acts like the film in scenario-one (figure~\ref{fig:regime_map}(a)), then it will produce humps and plugs as the secretion rate is increased. This correspondence, if true, would allow us to estimate the transition values of $I_s$ by simply replacing $I_c$ in the scenario-one regime map by $I_c^*+I_s$.

\begin{figure}
\centering
 \includegraphics[width=.88\textwidth]{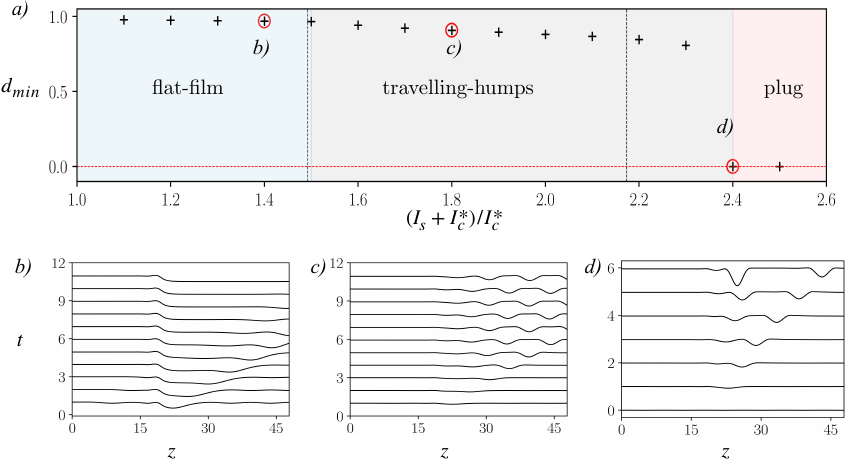}
\caption{\label{fig:secretion_flat_base} (\textit{a}) Regime map demarcating the different behaviours of the film downstream of the secretion zone (located at $z=z_s=19.7$); here, the mucus secretion rate $I_s$ is varied. Illustrations of the film's evolution in the three regimes are presented in panels (b-d). Here $u_c U = 60$ \si{\mu m.s^{-1}}; analogous results for $u_c U = 40$ \si{\mu m.s^{-1}} are presented in the \href{https://bighome.iitb.ac.in/index.php/s/yeijAoz47zg7yZk}{ supplementary material}.}  
\end{figure}

Figure~\ref{fig:secretion_flat_base}(a) presents the scenario-two regime map, constructed from simulations with secretion inside the domain and for the same value of $u_c$ as figure~\ref{fig:regime_map}(a); the predictions that follow from figure~\ref{fig:regime_map}(a), for the transition secretion-rates, 
are marked by vertical black dashed lines. The expected correspondence between the two scenarios is borne out, particularly for the onset of the travelling-humps regime. Plugging, however, occurs at slightly higher mucus-influx rates in the secretory scenario. Our simulations for $u_c U = 40$ 
\si{\mu m.s^{-1}} exhibit an analogous correspondence between the two scenarios (see the \href{https://bighome.iitb.ac.in/index.php/s/yeijAoz47zg7yZk}{ supplementary material}).

The film's evolution in the different regimes is illustrated in figures~\ref{fig:secretion_flat_base}(b-d) (see the animations in movie 4, movie 5, and movie 6 of the \href{https://bighome.iitb.ac.in/index.php/s/yeijAoz47zg7yZk}{ supplementary material}). As anticipated, the pre-secretion, upstream portion of the film is always flat because its thickness lies in the flat-film regime. The post-secretion, downstream portion of the film becomes flat at low secretion rates (figure~\ref{fig:secretion_flat_base}(b)) but develops travelling-humps at moderate secretion rates (figure~\ref{fig:secretion_flat_base}(c)). 
At even higher secretion rates, the downstream film forms a plug (figure~\ref{fig:secretion_flat_base}(d)). Interestingly, the plug forms near the secretion zone, akin to how the plug formed near the inlet in the non-secretory case (figure~\ref{fig:regime_map}(d)). We therefore have a general mechanism for the onset of plug formation in the presence of a constant input of mucus: increasing the input rate produces films of increasing thickness which are subject to increasingly strong capillary forces; eventually ciliary transport is unable to translate humps away from the input-zone; a hump then remains near the input-zone, accumulating mucus until its volume exceeds the maximum limit for a unduloid, after which it rapidly forms a plug. 

The onset of plugging in an open domain, with a constant input of mucus, occurs at a smaller base film thickness than that in a closed periodic domain. Periodic simulations with ciliary transport show that plugging is guaranteed to occur once $1-d_0^2 \gtrsim 1.73 R/\Lambda$ \citep{hazra2025probabilistic}, i.e., once $1-d_0 \gtrsim 0.117$ (this is when all the humps formed by the fastest-growing mode of the Rayleigh-Plateau instability exceed the maximum unduloid volume). The onset thickness reduces by approximately half in an open domain with a mucus source: plugging occurs in
figure~\ref{fig:regime_map}(a) once $1-d_0 \gtrsim 0.042$.

How does the plug evolve after it is formed? Will it grow in width while remaining in place, or will it be transported out of the airway into a larger downstream-airway where it might split apart? To answer this question, the WRIBL thin-film model would have to be augmented with the pseudo-plug treatment of \citep{Dietze2020occlusion}, which has recently been employed to study air-driven transport of liquid plugs \citep{Dietze2024plugs}. This is a promising direction for future work.

\section{Concluding remarks}

We have shown that ciliary transport alters the distribution of pulmonary mucus films, provided the open boundaries of the airway are taken into account. On the one hand, the Rayleigh-Plateau instability of thin films is rendered convective by ciliary transport, which thereby stabilizes the uniform flat-film state. On the other hand, the thickness required for plugs to form is reduced in the presence of a constant influx of mucus---whether by cilia-driven transport from an upstream airway or by direct secretion inside the airway.

The flat-film regime is crucial for lung health because it corresponds to a uniform coating of mucus that will protect the airway walls from inhaled allergens and pathogens. In contrast, the travelling-humps regime, with its mucus-depleted zones, leaves large portions of the airway wall vulnerable to particle-deposition \citep{hazra-particles}. 

The flat-film only emerges when one accounts for open boundaries and ciliary transport. The incorporation of these features in the present study is aided by the assumption of a viscous mucus film, the use of the WRIBL model, and the simple wall-velocity treatment of ciliary action. In the future, this minimal description should be systematically elaborated by accounting for non-Newtonian mucus-rheology \citep{Romano2023viscoplastic,Dietze2023mucociliary}, biosurfactants \citep{romano2022effect}, and two-way coupling between the motions of mucus and cilia; the latter calls for a multiscale framework that can extend the insights of small-scale, cilia-resolving simulations~\citep{smith2018Blaketribute,Sedaghat2016lbm} to the large scales of the airway.

 \vspace{.5\baselineskip}
 \noindent\small{\textbf{Acknowledgements.} {J.R.P. acknowledges his Associateship with the International Centre for Theoretical Sciences (ICTS), Tata Institute of Fundamental Research, Bangalore, India. The authors thank the National PARAM Supercomputing Facility \textit{PARAM SIDDHI-AI} at CDAC, Pune for computing resources; simulations were also performed on the IIT Bombay workstations \textit{Gandalf} (procured through DST-SERB grant SRG/2021/001185), and \textit{Faramir} and \textit{Aragorn} (procured through the IIT-B grant RD/0519-IRCCSH0-021).}


\vspace{.5\baselineskip}
\noindent\small{\textbf{Funding.} {This work was supported by
DST-SERB (J.R.P., grant no. SRG/2021/001185) and IRCC, IIT Bombay (S.H., Ph.D. fellowship; J.R.P., grant no. RD/0519-IRCCSH0-021).
}




\vspace{.5\baselineskip}
\noindent\small{\textbf{Author ORCHID.} {S.~Hazra, https://orcid.org/0009-0000-3728-5888; J.~R.~Picardo, https://orcid.org/0000-0002-9227-5516}}



\FloatBarrier

\bibliographystyle{jfm}
\bibliography{mucus}

@article{Dietze2015,
	Author = {Dietze, G. F. and Ruyer-Quil, C. },
	Journal = {J. Fluid Mech.},
	Pages = {68--109},
	Title = {Films in narrow tubes},
	Volume = {762},
	Year = {2015}}

@article{Dietze2023mucociliary, 
title={On the role of viscoelasticity in mucociliary clearance: a hydrodynamic continuum approach}, 
volume={971}, 
journal={J. Fluid Mech.}, 
author={Choudhury, A. and Filoche, M. and Ribe, N. M. and Grenier, N. and Dietze, G. F.}, 
year={2023}, 
pages={A33}
}

@article{Dietze2020occlusion, 
title={Falling liquid films in narrow tubes: occlusion scenarios}, 
volume={894}, 
journal={J. Fluid Mech.}, 
author={Dietze, G. F. and Lavalle, G. and Ruyer-Quil, C.}, 
year={2020}, 
pages={A17}}

@article{Dietze2024plugs, 
title={Liquid plugs in narrow tubes: application to airway occlusion}, 
volume={998},
journal={J. Fluid Mech.}, 
author={Dietze, G. F.}, 
year={2024}, pages={A50}}

@article{Bottier2017model,
    author = {Bottier, M. AND Peña Fernández, M. AND Pelle, G. AND Isabey, D. AND Louis, B. AND Grotberg, J. B. AND Filoche, M.},
    journal = {PLOS Comput. Biol.},
    title = {A new index for characterizing micro-bead motion in a flow induced by ciliary beating: Part II, modeling},
    year = {2017},
    month = {07},
    volume = {13},
    pages = {1-21},
    number = {7}
}

@article{romano2022effect,
  title={Effect of surfactant in an airway closure model},
  author={Roman{\`o}, F. and Muradoglu, M. and Grotberg, J. B.},
  journal={Phys. Rev. Fluids},
  volume={7},
  number={9},
  pages={093103},
  year={2022},
  publisher={APS}
}

@article{Levy2014,
	Author = {R. Levy and D. B. Hill and M. G. Forest and J. B. Grotberg},
	Journal = {Integr. Comp. Biol.},
	Pages = {985–1000},
	Title = {Pulmonary Fluid Flow Challenges for Experimental and Mathematical Modeling},
	Volume = {54},
	Year = {2014}}

@article{rogers2004airway,
  title={Airway mucus hypersecretion in asthma: an undervalued pathology?},
  author={Rogers, D. F.},
  journal={Curr. Opin. Pharmacol.},
  volume={4},
  number={3},
  pages={241--250},
  year={2004},
  publisher={Elsevier}
}

@article{button2012,
  title={A periciliary brush promotes the lung health by separating the mucus layer from airway epithelia},
  author={Button, B. and Cai, L. H. and Ehre, C. and Kesimer, M. and Hill, D. B. and Sheehan, J. K. and Boucher, R. C. and Rubinstein, M.},
  journal={Science},
  volume={337},
  number={6097},
  pages={937--941},
  year={2012},
  publisher={American Association for the Advancement of Science}
}

@article{Sedaghat2016lbm,
title = {Numerical simulation of muco-ciliary clearance: immersed boundary-lattice Boltzmann method},
journal = {Comput. Fluids},
volume = {131},
pages = {91-101},
year = {2016},
author = {M. H. Sedaghat and M. M. Shahmardan and M. Norouzi and P. G. Jayathilake and M. Nazari}
}

@article{vasquez2016modeling,
  title={Modeling and simulation of mucus flow in human bronchial epithelial cell cultures--Part I: Idealized axisymmetric swirling flow},
  author={Vasquez, P. A. and Jin, Y. and Palmer, E. and Hill, D. and Forest, M. G.},
  journal={PLoS Comput. Biol.},
  volume={12},
  number={8},
  pages={e1004872},
  year={2016},
  publisher={Public Library of Science San Francisco, CA USA}
}

@article{rayleigh1892xvi,
  title={On the instability of a cylinder of viscous liquid under capillary force},
  author={Rayleigh},
  journal={The London, Edinburgh, and Dublin Philosophical Magazine and Journal of Science},
  volume={34},
  number={207},
  pages={145--154},
  year={1892},
  publisher={Taylor \& Francis}
}

@article{lister2006capillary,
  title={Capillary drainage of an annular film: the dynamics of collars and lobes},
  author={Lister, J. R. and Rallison, J. M. and King, A. A. and Cummings, L. J. and Jensen, O. E.},
  journal={J. Fluid Mech.},
  volume={552},
  pages={311--343},
  year={2006},
  publisher={Cambridge University Press}
}

@article{heil2008mechanics,
  title={The mechanics of airway closure},
  author={Heil, M. and Hazel, A. L. and Smith, J. A.},
  journal={Respir. Physiol. Neurobiol.},
  volume={163},
  number={1-3},
  pages={214--221},
  year={2008},
  publisher={Elsevier}
}

@article{everett1972model,
  title={Model studies of capillary condensation. I. Cylindrical pore model with zero contact angle},
  author={Everett, D. H. and Haynes, J. M.},
  journal={J. Colloid Interface Sci.},
  volume={38},
  number={1},
  pages={125--137},
  year={1972},
  publisher={Elsevier}
}

@article{smith2018Blaketribute, 
title={Biological fluid mechanics under the microscope: a tribute to John Blake}, volume={59},
number={4}, 
journal={ANZIAM J.}, 
author={Smith, D.J.}, 
year={2018}, 
pages={416–442}}

@article{Romano2019viscous,
  title = {Liquid plug formation in an airway closure model},
  author = {Roman{\`o}, F. and Fujioka, H. and Muradoglu, M. and Grotberg, J. B.},
  journal = {Phys. Rev. Fluids},
  volume = {4},
  issue = {9},
  pages = {093103},
  numpages = {23},
  year = {2019}
}

@article{Romano2023viscoplastic,
  title = {Effects of elastoviscoplastic properties of mucus on airway closure in healthy and pathological conditions},
  author = {Erken, O. and Fazla, B. and Muradoglu, M. and Izbassarov, D. and Roman{\`o}, F. and Grotberg, J. B.},
  journal = {Phys. Rev. Fluids},
  volume = {8},
  issue = {5},
  pages = {053102},
  numpages = {26},
  year = {2023}
}

@article{pedley1977pulmonary,
  title={Pulmonary fluid dynamics},
  author={Pedley, T. J.},
  journal={Annu. Rev. Fluid Mech.},
  volume={9},
  number={1},
  pages={229--274},
  year={1977},
  publisher={Annual Reviews 4139 El Camino Way, PO Box 10139, Palo Alto, CA 94303-0139, USA}
}

@article{Boucher2006,
author = {S. H. Randell AND R. C. Boucher},
journal = {Am. J. Respir. Cell Mol. Biol. },
title = {Effective mucus clearance is essential for respiratory health},
year = {2006},
volume = {35},
pages = {20-28},
number = {1}
}

@article{Sleigh1988,
author = {Sleigh, M. A. and Blake, J. R. and Liron, N.},
journal = {Am Rev. Respir. Dis. },
title = {The propulsion of mucus by cilia},
year = {1988},
volume = {137},
pages = {726–741},
number = {3}
}

@article{scipy,
  author  = {Virtanen, P. and Gommers, R. and Oliphant, T. E. and Haberland, M. and Reddy, T. and Cournapeau, D. and Burovski, E. and Peterson, P. and Weckesser, W. and Bright, J. and others and {SciPy 1.0 Contributors}},
  title   = {{{SciPy} 1.0: Fundamental Algorithms for Scientific
            Computing in Python}},
  journal = {Nat. Methods},
  year    = {2020},
  volume  = {17},
  pages   = {261--272},
comment = {solve{\_}ivp documentation: https://docs.scipy.org/doc/scipy/reference/generated/scipy.integrate.solve{\_}ivp.html}
}

@article{oron1997long,
  title={Long-scale evolution of thin liquid films},
  author={Oron, A. and Davis, S. H. and Bankoff, S. G.},
  journal={Rev. Mod. Phys.},
  volume={69},
  number={3},
  pages={931},
  year={1997},
  publisher={APS}
}

@unpublished{hazra-particles,
title={Aerosol deposition in mucus-lined ciliated airways}, 
author={S. Hazra and J. R. Picardo},
year={2025},
note = {\texttt{arXiv:2502.01883}}
}

@article{johnson1991,
  title={The nonlinear growth of surface-tension-driven instabilities of a thin annular film},
  author={Johnson, M. and Kamm, R. D. and Ho, L. W. and Shapiro, A. and Pedley, T. J.},
  journal={J. Fluid Mech.},
  volume={233},
  pages={141--156},
  year={1991}
}

@article{hazra2025probabilistic,
  title = {Probabilistic plugging of airways by sliding mucus films},
  author = {Hazra, S. and Picardo, J. R.},
  journal = {Phys. Rev. Fluids},
  volume = {10},
  pages={094002},
  year = {2025}
}

@article{Dickey2018-secretion,
author = {Jaramillo, A. M. and Azzegagh, Z. and Tuvim, M. J. and Dickey, B. F.},
title = {Airway Mucin Secretion},
journal = {Annals of the American Thoracic Society},
volume = {15},
number = {Supplement\_3},
pages = {S164-S170},
year = {2018},
doi = {10.1513/AnnalsATS.201806-371AW},
URL = {https://doi.org/10.1513/AnnalsATS.201806-371AW}
}

@article{Rogers2033-goblet,
title = {The airway goblet cell},
journal = {Int. J. Biochem. Cell Biol.},
volume = {35},
number = {1},
pages = {1-6},
year = {2003},
issn = {1357-2725},
doi = {https://doi.org/10.1016/S1357-2725(02)00083-3},
url = {https://www.sciencedirect.com/science/article/pii/S1357272502000833},
author = {Duncan F Rogers}
}

@article{William2015-baseline,
    doi = {10.1371/journal.pone.0127267},
    author = {Zhu, Y. AND Abdullah, L. H. AND Doyle, S. P. AND Nguyen, K. AND Ribeiro, C. M. P. AND Vasquez, P. A. AND Forest, M. G. AND Lethem, M. I. AND Dickey, B. F. AND Davis, C. W.},
    journal = {PLOS ONE},
    title = {Baseline Goblet Cell Mucin Secretion in the Airways Exceeds Stimulated Secretion over Extended Time Periods, and Is Sensitive to Shear Stress and Intracellular Mucin Stores},
    year = {2015},
    month = {05},
    volume = {10},
    url = {https://doi.org/10.1371/journal.pone.0127267},
    pages = {1-27},
    number = {5}
}

@article{rees2010effect,
  title={The effect of confinement on the stability of viscous planar jets and wakes},
  author={Rees, S. J. and Juniper, M. P.},
  journal={Journal of Fluid Mechanics},
  volume={656},
  pages={309--336},
  year={2010},
  publisher={Cambridge University Press}
}

@article{HazraWRIBL,
title={Deriving thin-film averaged equations using computer algebra}, 
author={S. Hazra and J. R. Picardo},
year={2025},
  journal={J. Eng. Math.},
  volume={154},
  pages={5}
}

@Article{Numpy2020,
 title         = {Array programming with {NumPy}},
 author        = {C. R. Harris and K. J. Millman and S. J.
                 van der Walt and R. Gommers and P. Virtanen and D.
                 Cournapeau and Eric Wieser and Julian Taylor and Sebastian
                 Berg and Nathaniel J. Smith and Robert Kern and Matti Picus
                 and Stephan Hoyer and Marten H. van Kerkwijk and Matthew
                 Brett and Allan Haldane and Jaime Fern{\'{a}}ndez del
                 R{\'{i}}o and Mark Wiebe and Pearu Peterson and Pierre
                 G{\'{e}}rard-Marchant and Kevin Sheppard and Tyler Reddy and
                 Warren Weckesser and Hameer Abbasi and Christoph Gohlke and
                 Travis E. Oliphant},
 year          = {2020},
 month         = sep,
 journal       = {Nature},
 volume        = {585},
 number        = {7825},
 pages         = {357--362},
 doi           = {10.1038/s41586-020-2649-2},
 url           = {https://doi.org/10.1038/s41586-020-2649-2}
}

@incollection{Huerre2000Shear,
  author    = {P. Huerre},
  title     = {Open Shear Flow Instabilities},
  booktitle = {Perspectives in Fluid Dynamics: A Collective Introduction to Current Research},
  editor    = {G. K. Batchelor and H. K. Moffatt and M. G. Worster},
  publisher = {Cambridge University Press},
  address   = {Cambridge},
  year      = {2000},
  pages     = {159--229}
}

\end{document}